\documentclass{webofc}
\usepackage[varg]{txfonts}   
\usepackage{caption}
\usepackage{subcaption}

\begin{document}
\newcommand{\nue}{\nu_\textrm{e}}
\title{MicroBooNE Public Data Sets: a Collaborative Tool for LArTPC Software Development}
%
%

\author{\firstname{Giuseppe} \lastname{Cerati}\inst{1}, for the MicroBooNE Collaboration\thanks{\email{MICROBOONE_INFO@fnal.gov}}
}

\institute{Fermi National Accelerator Laboratory, Kirk and, Pine St Batavia, IL 60510}

\abstract{%
  Among liquid argon time projection chamber (LArTPC) experiments MicroBooNE is the one that continually took physics data for the longest time (2015-2021), and represents the state of the art for reconstruction and analysis with this detector. Recently published analyses include oscillation physics results, searches for anomalies and other BSM signatures, and cross section measurements. LArTPC detectors are being used in current experiments such as ICARUS and SBND, and being planned for future experiments such as DUNE. MicroBooNE has recently released to the public two of its data sets, with the goal of enabling collaborative software developments with other LArTPC experiments and with AI or computing experts. These data sets simulate neutrino interactions on top of off-beam data, which include cosmic ray background and noise. The data sets are released in two formats: the native art/ROOT format used internally by the collaboration and familiar to other LArTPC experts, and the HDF5 format which contains reduced and simplified content and is suitable for usage by the broader community. This contribution presents the open data sets, discusses their motivation, the technical implementation, and the extensive documentation -- all inspired by FAIR principles. Finally, opportunities for collaborations are discussed.
}
\maketitle
%

\section{Introduction}
\label{intro}
MicroBooNE~\cite{MicroBooNE:2016pwy} is a neutrino experiment at Fermilab designed to test the MiniBooNE anomaly~\cite{MiniBooNE:2020pnu} as it is located along the same Booster Neutrino Beam (BNB) and at a similar distance from source. MicroBooNE's goals are not limited to probing this anomaly, and span a broader experimental program including tests of short-baseline oscillations as part of the Short Baseline Neutrino program (SBN)~\cite{MicroBooNE:2015bmn}, searches for beyond-Standard Model particles, and the measurement of neutrino-argon interaction cross sections.
MicroBooNE's physics operations took place between 2015 and 2021. To this date, MicroBooNE has analyzed about half of the collected beam data and produced over 50 publications.

MicroBooNE's detector~\cite{MicroBooNE:2016pwy} is a liquid argon time projection chamber (LArTPC). The working principle of the MicroBooNE LArTPC is the following: charged particles produced in neutrino-argon interactions ionize the argon as they travel in the detector active volume (2.56 $\times$ 2.32 $\times$ 10.36 m). Ionization electrons drift in an electric field towards anode planes. Here, sense wires detect the incoming charge. About 8200 wires are arranged in three planes oriented in different directions (0, $\pm$60 degrees), allowing for 3D reconstruction of the particle trajectories with $\mathcal{O}$(mm) spatial resolution. The amplitude of the signal detected on the wires provides calorimetric information for energy measurements. The fast scintillation light emitted by the argon is detected by the optical system, made of 32 photo-multiplier tubes (PMTs), and is used for triggering and cosmic rejection.

In this document we describe the release of MicroBoNE data sets for the purpose of collaborative software development. This release is motivated by the following arguments:
\begin{itemize}
    \item Establish MicroBooNE as state of the art LArTPC technology. This is already attested by our publication record, but public data sets provide a direct reference point for any LArTPC software development.
    \item Efficient collaboration of members of MicroBooNE with colleagues in other LArTPC experiments, as well as with computer scientists. Until this data release, software development collaborations were required to have an approved memorandum of understanding to share data sets outside the Collaboration or to use other public data sets. Being able to use MicroBooNE data sets implies that the output of external collaborations is directly usable within MicroBooNE without further tuning.
    \item Potentially attract developments from beyond our community, through public initiatives such as data challenges.
\end{itemize}

The next sections describe the technical implementation of the data release, as well as the relative documentation. Finally, conclusions and future prospects are discussed.

\section{Implementation of open samples}

With this data release, MicroBooNE aims at reaching out to the largest possible set of developers and at enabling the widest range of applications. In the process, we therefore followed as much as possible the FAIR principles for scientific data management (findable, accessible, interoperable, reusable data). For more information on these principles, see e.g. ref.~\cite{Chen:2021euv}. The MicroBooNE open samples are advertized from the MicroBooNE website\footnote{https://microboone.fnal.gov/documents-publications/public-datasets/}, and are made available on the Zenodo open data repository. The MicroBooNE website contains a brief description of the data set, links to Zenodo and to documentation, and information about license and citation. Zenodo, provides citable DOI (digital object identifier) and versioning. Samples are made available under the "cc-by" license, allowing users to utilize the data in any way, including modifying and redistributing, as long as credit is given to the original authors. A suggested text for acknowledging the Collaboration is provided. The Collaboration requests that, whenever possible, resulting software products are also made publicly available, although this is not required by the license. 

The data sets release consists of "overlay" samples, where with overlay we refer to events  from off-beam data taking with an overlaid simulated neutrino interaction. These events provide data-driven cosmic ray background and noise, as well as Monte Carlo truth information for the neutrino interaction. Neutrino interactions in the open samples are either the inclusive set of neutrino interactions as expected from the BNB nominal flux or the subset of charged-current electron neutrino interactions in the BNB. We will refer to these as "inclusive BNB" and "intrinsic $\nue$", respectively. Inclusive BNB interactions are simulated in the full cryostat volume, while intrinsic $\nue$ interactions are simulated in the LArTPC active volume.

The data is released in two formats. The first one is the art/ROOT data format~\cite{Green:2012gv,Brun:1997pa}. These are the same files used within the collaboration, thus making available to the public all reconstructed and simulated data products. This data format targets HEP physicists, and in particular the LArTPC community, that are likely already familiar with this data format. art/ROOT files are stored on a dedicated persistent dCache pool area that is accessible with xrootd~\cite{xrootd} without requiring any virtual organization credentials. The list of xrootd urls is stored on Zenodo. 
The second format is HDF5~\cite{Lee:2021uhu}, targeting usage by the broader data and computer science communities. HDF5 files include a subset of the art/ROOT information, with a simplified layout for ease of use. Nevertheless, this data format contains the most useful information and is designed to allow a wide range applications. The following information is stored in the HDF5 files:
\begin{enumerate}
    \item Noise-filtered and deconvolved wire waveforms in regions of interest.
    \item LArTPC hit information.
    \item Optical hit and flash information.
    \item Monte Carlo truth information (incoming neutrino properties, energy deposits as associated to hits, Geant4~\cite{GEANT4:2002zbu} particles).
\end{enumerate}
In addition we provide information for the purpose of benchmarking new developments against the state-of-the-art reconstruction performance provided by the Pandora multi-algorithm pattern recognition toolkit~\cite{MicroBooNE:2017xvs}. These include the interaction and cluster hit mapping, a multivariate track-shower classification, the neutrino flavor identification.
HDF5 files stored on Zenodo, at the same DOI as the xrootd urls of corresponding art/ROOT data set. Each HDF5 sample comes in two flavors: with and without wire waveform information listed as 1. in the list above. Due to size requirements, samples with this information contain less events. A summary of the samples can be found in table~\ref{tab:samples}.

\begin{table}
\centering
\caption{Summary of MicroBooNE Open Data Sets. The column "Wire" refers to whether the wire waveform information is stored in the HDF5 files or not.}
\label{tab:samples}
\resizebox{\linewidth}{!}{
\begin{tabular}{ll|llll|lll}
\hline
Interaction & DOI & \multicolumn{4}{c}{HDF5} & \multicolumn{3}{|c}{art/ROOT} \\
& & Wire & Events & Files & Size & Events & Files & Size \\
\hline
Inclusive BNB & \cite{abratenko_2023_8370883} & No & 753,467 & 18 & 195 GB & 1,046,139 & 24,436 & 6.4 TB \\
Inclusive BNB & \cite{abratenko_2022_7262009} & Yes & 24,332 & 18 & 44 GB & 24,332 & 720 & 136 GB \\
Intrinsic $\nue$ & \cite{abratenko_2022_7261921} & No & 89,339 & 20 & 31 GB & 89,339 & 2,151 & 761 GB \\
Intrinsic $\nue$ & \cite{abratenko_2022_7262140} & Yes & 19,940 & 20 & 39 GB & 19,940 & 540 & 170 GB \\
\hline
\end{tabular}
}
\end{table}

\section{Documentation}

Documentation is of utmost importance for an open data release, as it allows external users to become familiar with the content of the data and to understand which applications it can be used for. The art/ROOT format targets users from the LArTPC community, i.e. physicists already familiar with the LArSoft~\cite{Snider:2017wjd} software environment. Therefore, the documentation for this format assumes prior knowledge of LArSoft-related tools, and consists of:
\begin{itemize}
    \item A description of the samples and list of data products stored\footnote{https://github.com/uboone/OpenSamples/blob/v01/file-content-artroot.md}.
    \item Links to websites with documentation about the related software tools (LArSoft, xrootd, etc.).
    \item A recipe to setup the software release (uboonecode and LArSoft) from CVMFS.
    \item A link to the module for creating HDF5 files, providing an example of how to access the art/ROOT content.
\end{itemize}

Documentation for the HDF5 data sets instead needs to be more detailed, as these format targets users not necessarily familiar with LArTPC experiments and the software they use. We chose to document the usage of this format mainly through a demonstration of usage with a few Jupyter notebooks\footnote{https://github.com/uboone/OpenSamples/tree/v01}. These notebooks are described in the subsection below. A recipe for installing the required packages in a Conda environment to run the notebooks is provided. This environment has minimal dependencies so that users can easily install it and add any other package they may need for their applications on top. Dependencies include pynuml\footnote{https://github.com/vhewes/pynuml} for file I/O handling. All notebooks contain a brief introduction to clarify their purpose. We also provide a set of auxiliary tools, such as functions for basic detector navigation and minimal plotting utilities. Documentation also includes a description of the file content, formatted as a table with a brief explanation of each element stored in the data set.

\subsection{Notebooks Demonstrating Usage of HDF5 Files}

\begin{figure}[h]
\centering
\begin{subfigure}[b]{0.33\textwidth}
   \centering
   \includegraphics[width=\textwidth]{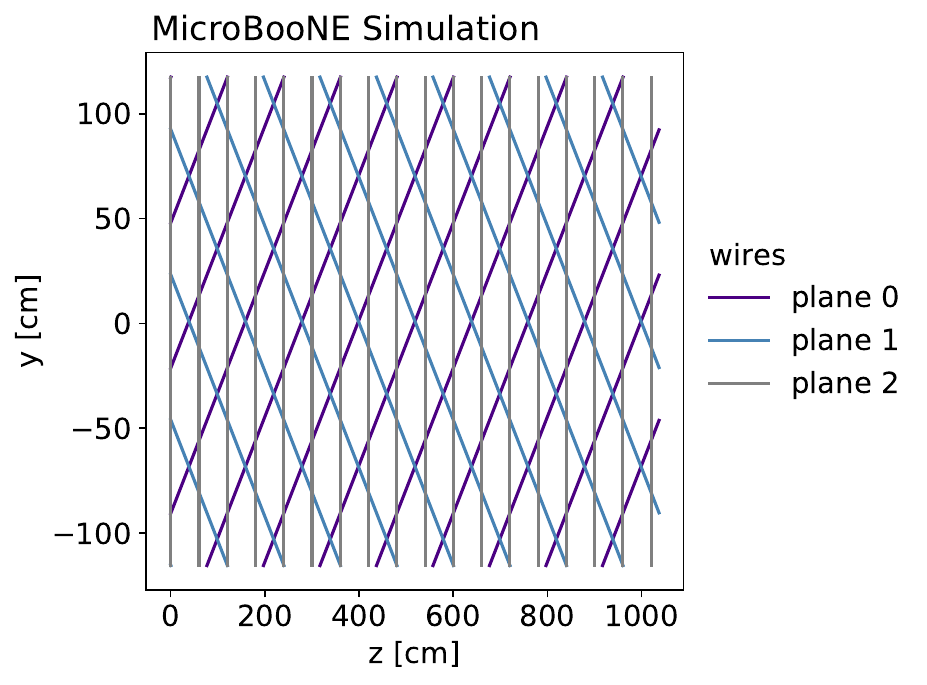}
   \caption{}
   \label{fig:expl_wires}
\end{subfigure}\hfill
\begin{subfigure}[b]{0.33\textwidth}
   \centering
   \includegraphics[width=\textwidth]{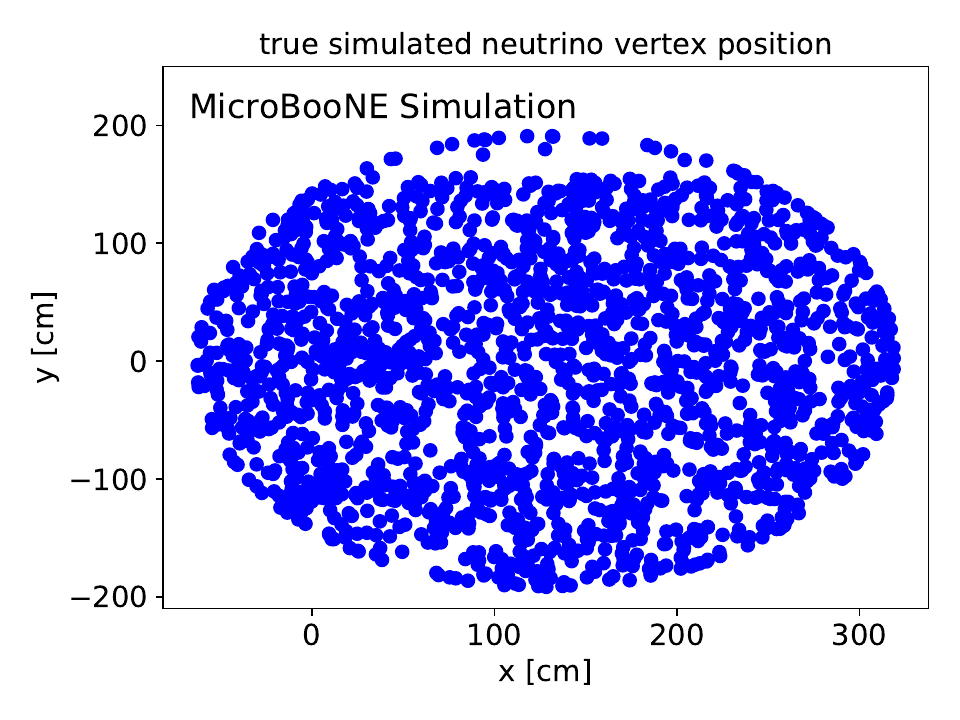}
   \caption{}
   \label{fig:expl_cryo}
\end{subfigure}\hfill
\begin{subfigure}[b]{0.33\textwidth}
   \centering
   \includegraphics[width=\textwidth]{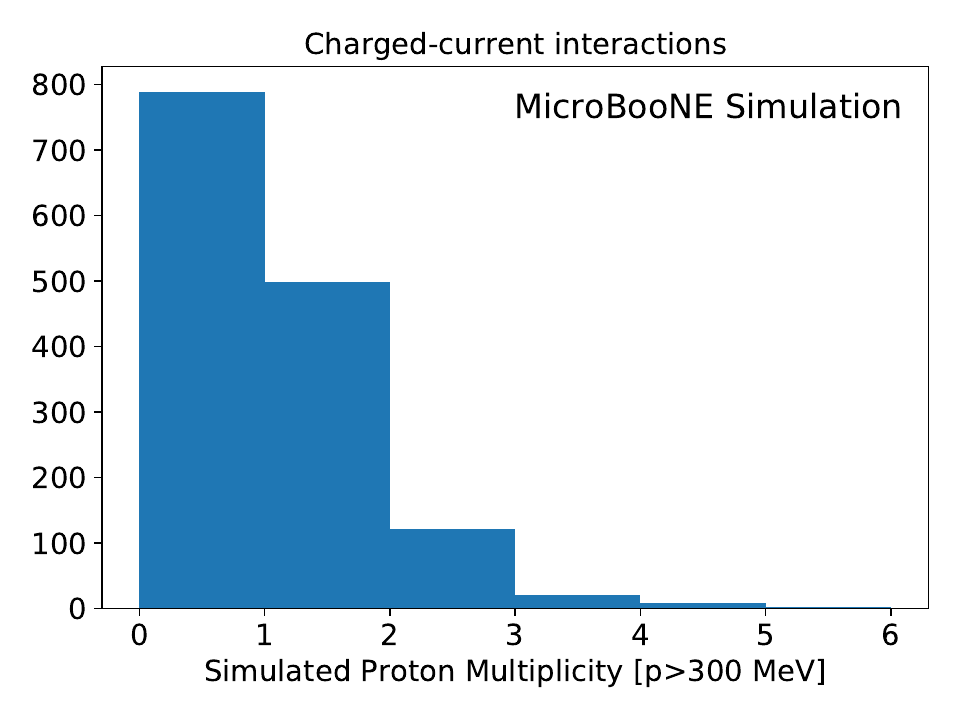}
   \caption{}
   \label{fig:expl_pmult}
\end{subfigure}\\
\caption{Example of plots from the "Sample Exploration" notebook: the detector positions of wires in each plane, displaying one wire every 200 (\subref{fig:expl_wires}), position of the simulated neutrino interaction vertex in the plane transverse to the beam (\subref{fig:expl_cryo}), number of protons with momentum above 300 MeV produced in BNB neutrino interactions (\subref{fig:expl_pmult}).}
\label{fig:expl}
\end{figure}

The first notebook, "Sample Exploration", is meant to help the user become familiar with the sample content and with tools provided to understand the detector properties.
As shown in figure~\ref{fig:expl}, the notebook demonstrates how to determine the wire positions and their intersections, as well properties of the neutrino interaction like the interaction position (vertex) in the cryostat and multiplicities of simulated particles.

\begin{figure}[h]
\centering
\begin{subfigure}[b]{\textwidth}
   \centering
   \includegraphics[width=0.75\textwidth]{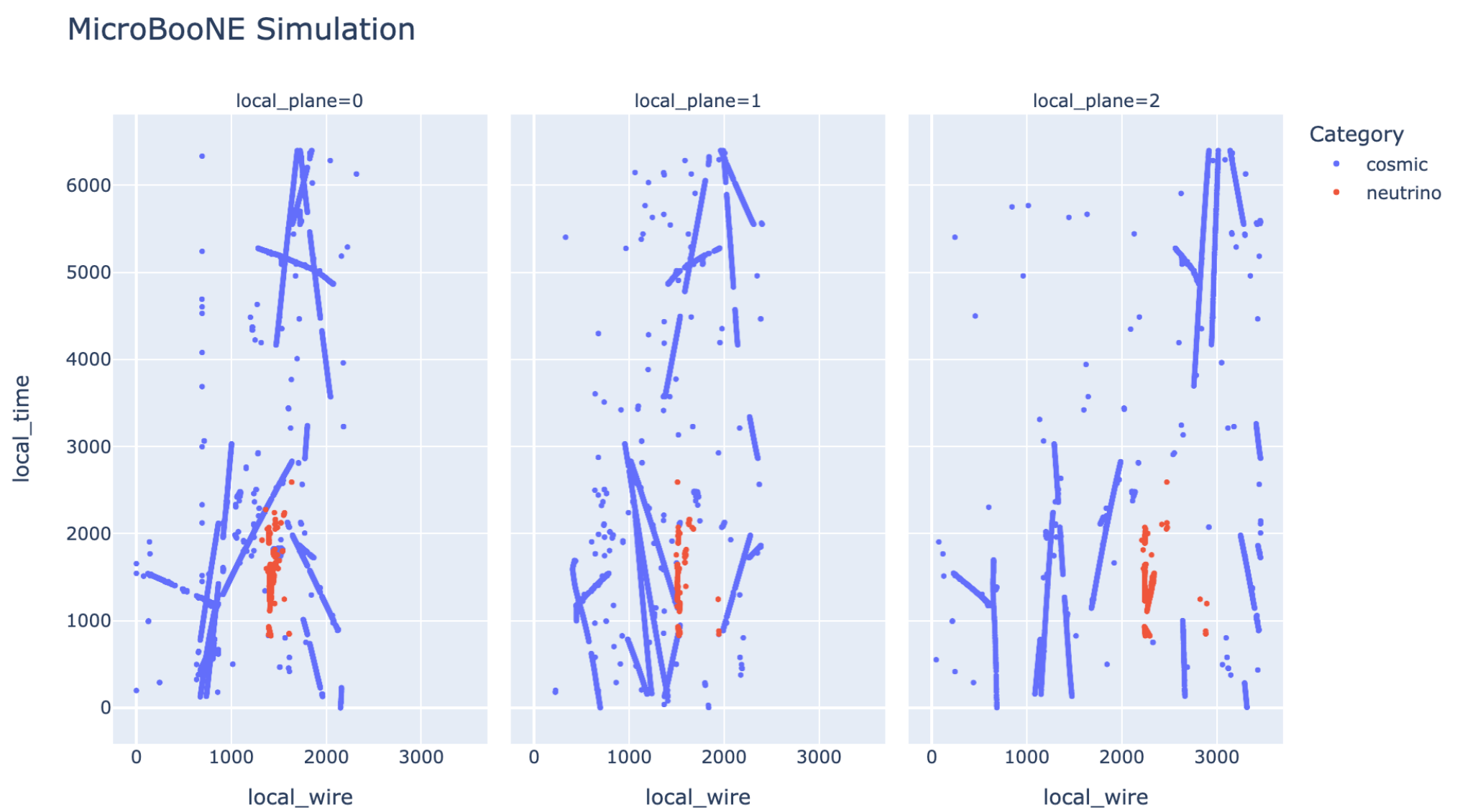}
   \caption{}
   \label{fig:hitl_cosm}
\end{subfigure}\\
\begin{subfigure}[b]{\textwidth}
   \centering
   \includegraphics[width=0.75\textwidth]{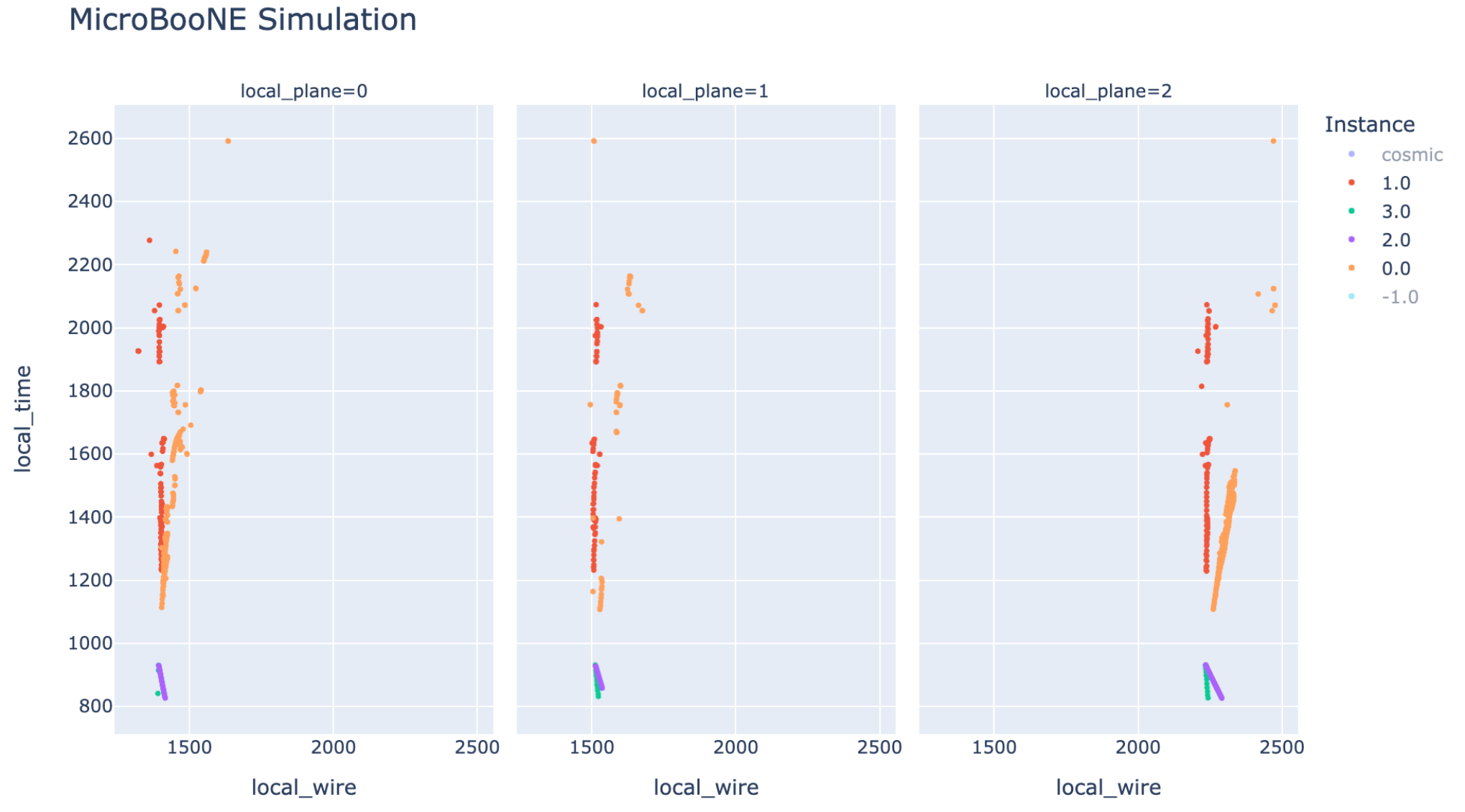}
   \caption{}
   \label{fig:hitl_inst}
\end{subfigure}
\caption{Example of plots from the "Hit Labeling" notebook: categorization of hits in the 3 planes based on whether they originate from a neutrino or cosmic ray interaction (\subref{fig:hitl_cosm}), categorization of neutrino hits in different particle instances (\subref{fig:hitl_inst}).}
\label{fig:hitl}
\end{figure}

The second notebook, "Hit Labeling", provides examples of ground-truth labels of TPC hits according to different categorizations. Each categorization can be the target of specific algorithms or network training. These categorizations are the neutrino identification (hits from neutrino interaction vs from noise or cosmic ray background), semantic segmentation (categorization of hits based on the type of particle that produced them), and instance segmentation (labeling of hits according to the particle instance that produced them). Examples are shown in figure~\ref{fig:hitl}.

\begin{figure}[h]
\centering
\begin{subfigure}[b]{\textwidth}
   \centering
   \includegraphics[width=0.75\textwidth]{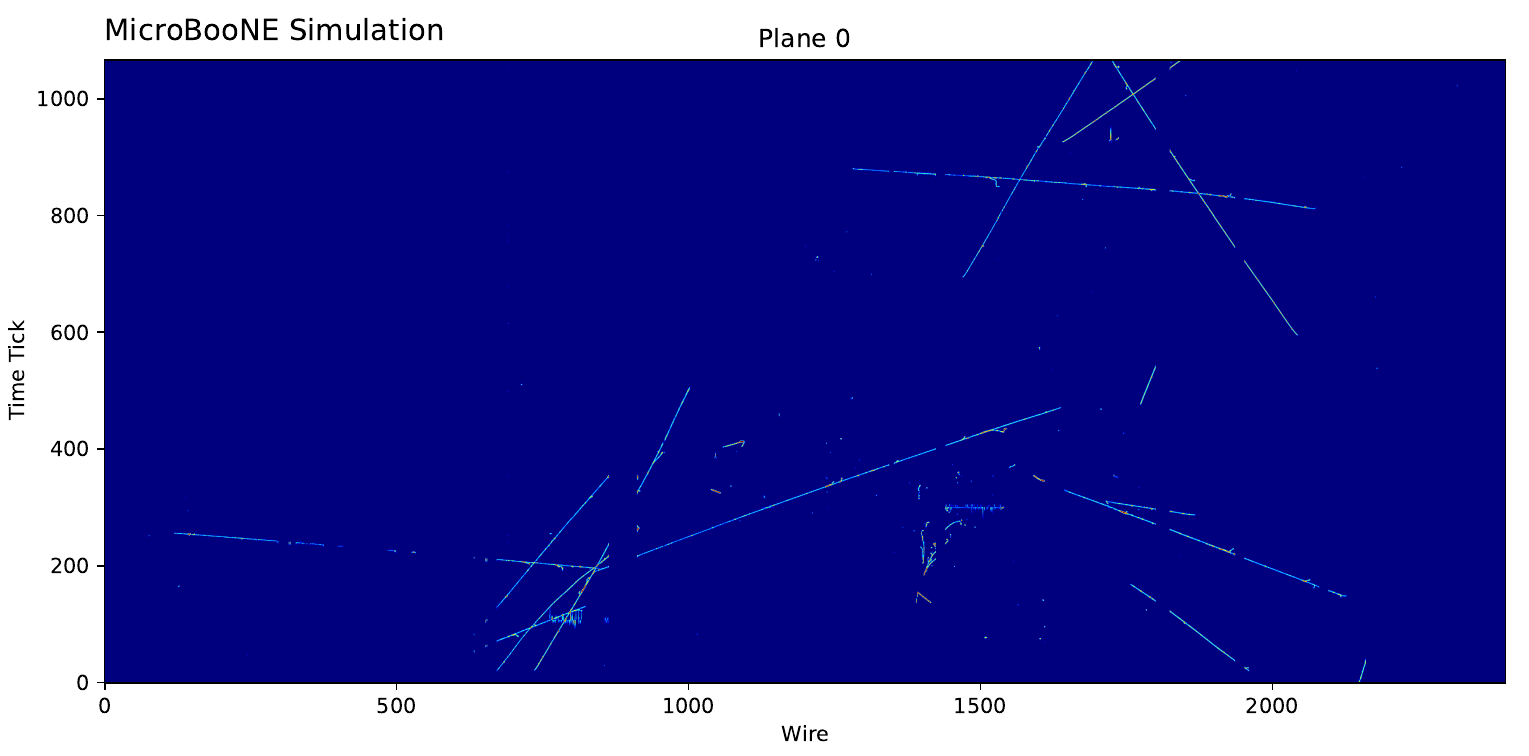}
   \caption{}
   \label{fig:wire_reco}
\end{subfigure}\\
\begin{subfigure}[b]{\textwidth}
   \centering
   \includegraphics[width=0.75\textwidth]{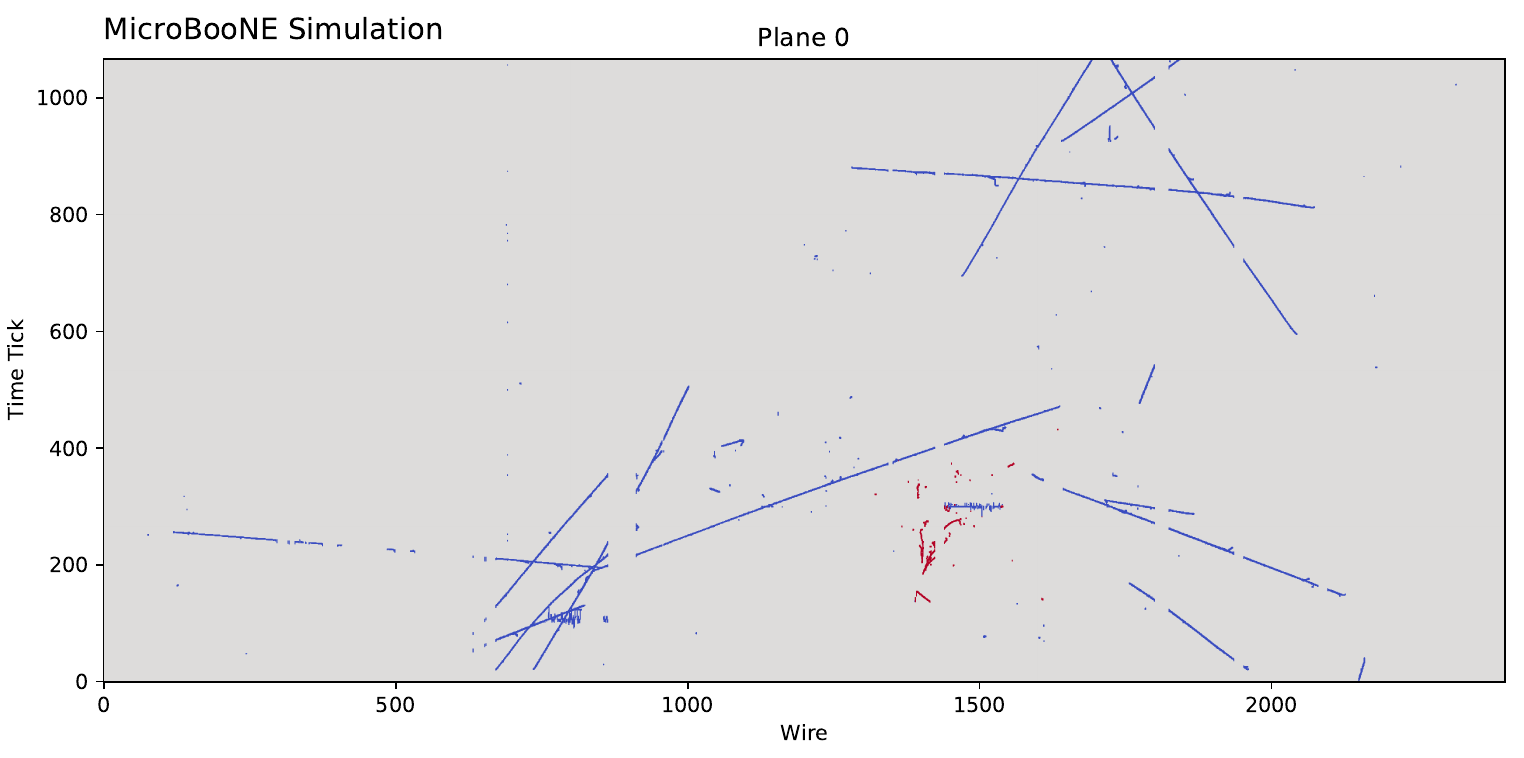}
   \caption{}
   \label{fig:wire_truth}
\end{subfigure}
\caption{Example of plots from the "WireImage" notebook: for the same event as in figure~\ref{fig:hitl}, the wire waveform image for plane 0 (\subref{fig:wire_reco}) and the corresponding ground truth in terms of neutrino identification (\subref{fig:wire_truth}) are shown.}
\label{fig:wire}
\end{figure}

The "WireImage" notebook is the only one that requires HDF5 files with the wire waveform information. This notebook demonstrates the LArTPC data visualization in image format. It can be used for visual data processing methods, such as Convolutional Neural Networks (CNN). Examples of CNNs developed by the MicroBooNE Collaboration can be found at refs~\cite{MicroBooNE:2020yze,MicroBooNE:2020hho,MicroBooNE:2018kka,MicroBooNE:2016dpb}.
Ground truth at waveform level is not directly provided, but the notebook demonstrates how it can be extracted by matching the waveform with the hit-level ground truth information. Example images of the wire waveform and of the extracted ground truth for a specific event are shown in figure~\ref{fig:wire}.

\begin{figure}[h]
\centering
\begin{subfigure}[b]{0.33\textwidth}
   \centering
   \includegraphics[width=\textwidth]{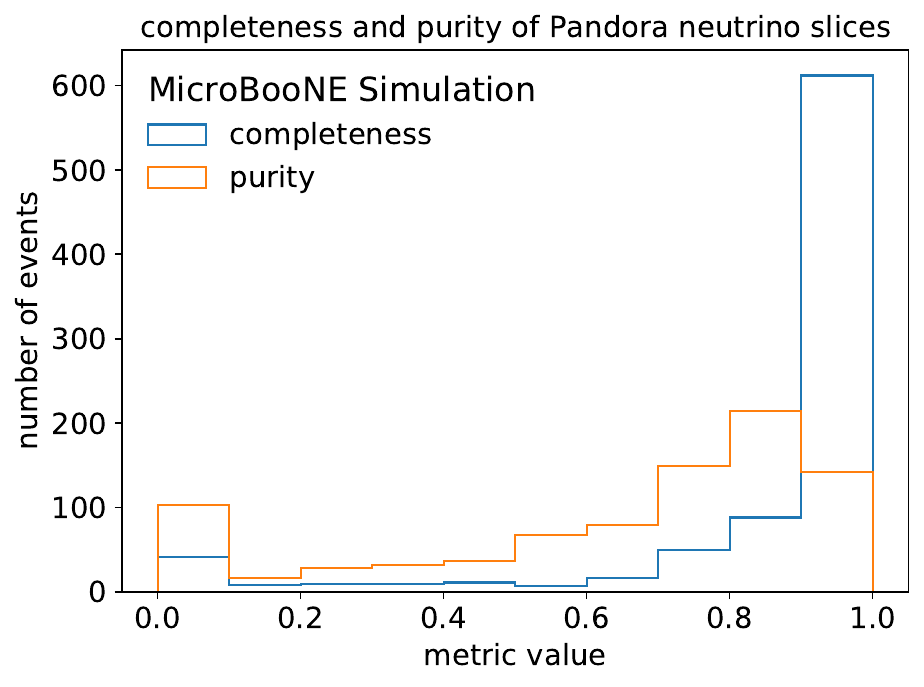}
   \caption{}
   \label{fig:pndr_slices}
\end{subfigure}\hfill
\begin{subfigure}[b]{0.33\textwidth}
   \centering
   \includegraphics[width=\textwidth]{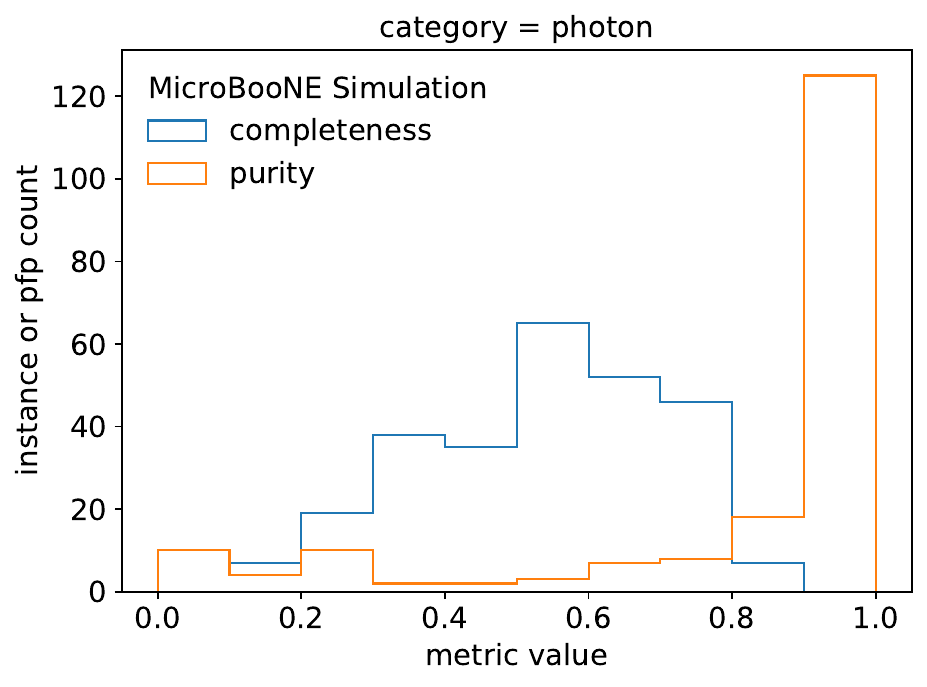}
   \caption{}
   \label{fig:pndr_instnc}
\end{subfigure}\hfill
\begin{subfigure}[b]{0.33\textwidth}
   \centering
   \includegraphics[width=\textwidth]{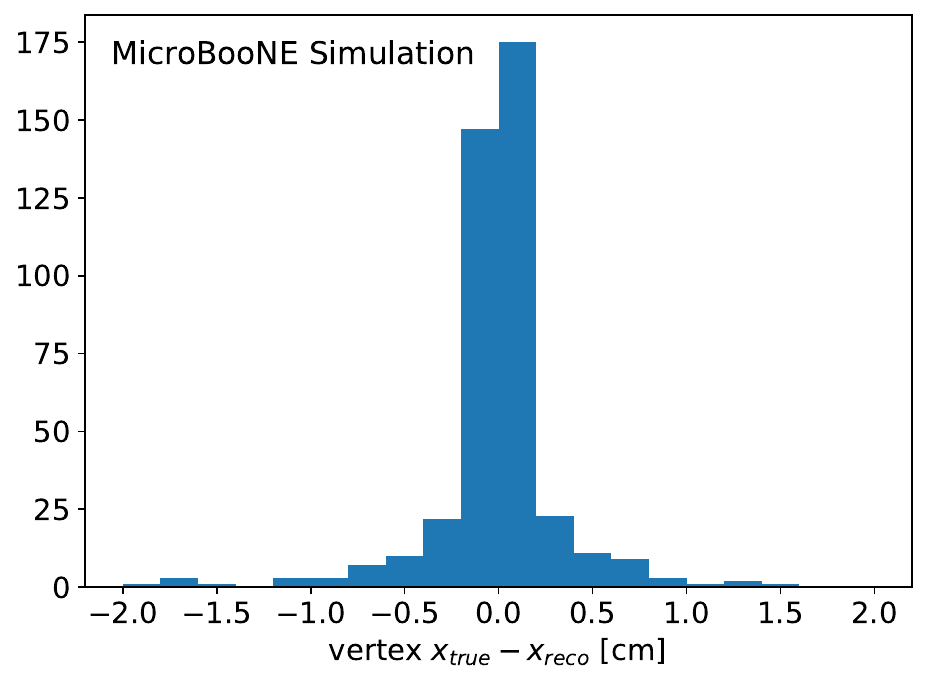}
   \caption{}
   \label{fig:pndr_vtxres}
\end{subfigure}\\
\caption{Example of plots from the "Pandora metrics" notebook: the purity and completeness for the neutrino interaction identification (\subref{fig:pndr_slices}) and for the reconstruction and clustering of hits from photons (\subref{fig:pndr_instnc}), resolution for the neutrino vertex $x$ coordinate (\subref{fig:pndr_vtxres}).}
\label{fig:pndr}
\end{figure}

Purpose of the "Pandora metrics" notebook is to introduce the definition of important
metrics used to benchmark the reconstruction software performance, and produce results for these metrics using Pandora. As shown in figure~\ref{fig:pndr}, these metrics include the neutrino vertex position resolution, as well as purity and completeness at the level of the neutrino interaction or at the level of particle instances. Purity and completeness are respectively defined as the fraction of reconstructed or true hits correctly identified.

\begin{figure}[h!]
\centering
\begin{subfigure}[b]{0.4\textwidth}
   \centering
   \includegraphics[width=\textwidth]{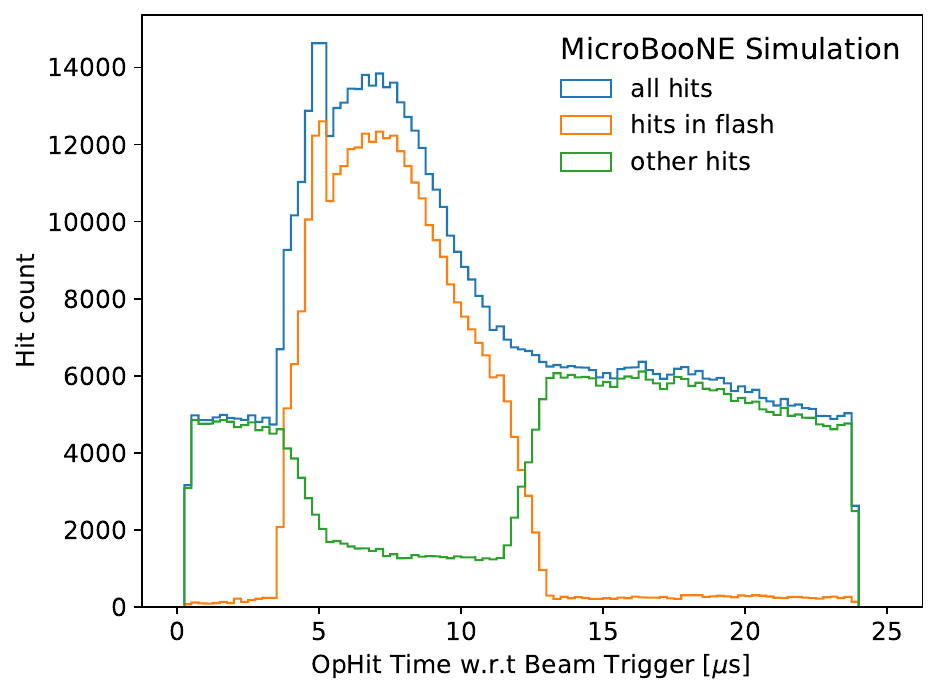}
   \caption{}
   \label{fig:optic_time}
\end{subfigure}\hfill
\begin{subfigure}[b]{0.4\textwidth}
   \centering
   \includegraphics[width=\textwidth]{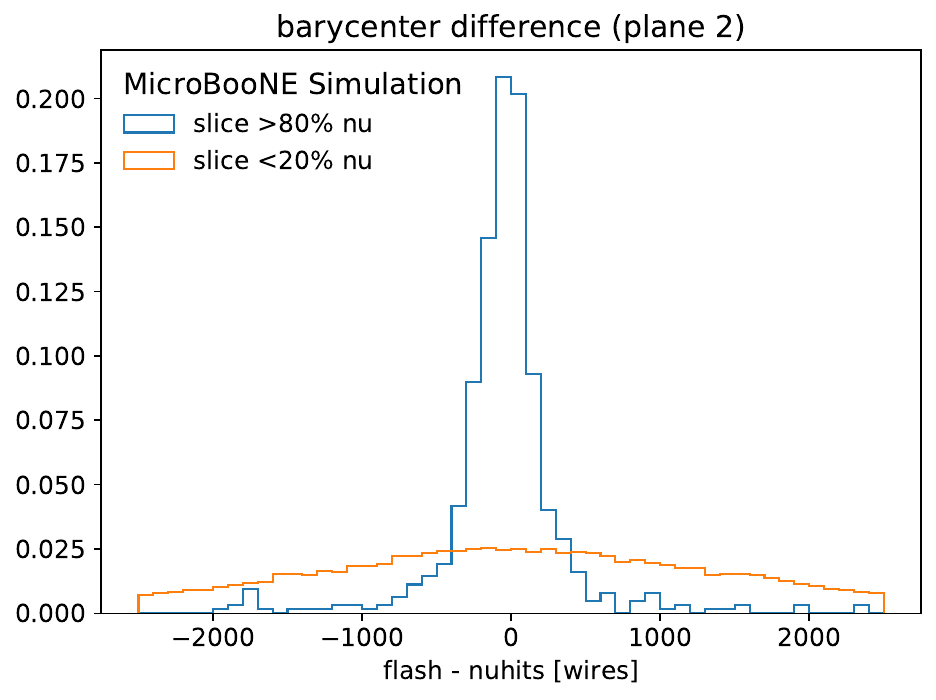}
   \caption{}
   \label{fig:optic_baryc}
\end{subfigure}\\
\begin{subfigure}[b]{0.6\textwidth}
   \centering
   \includegraphics[width=\textwidth]{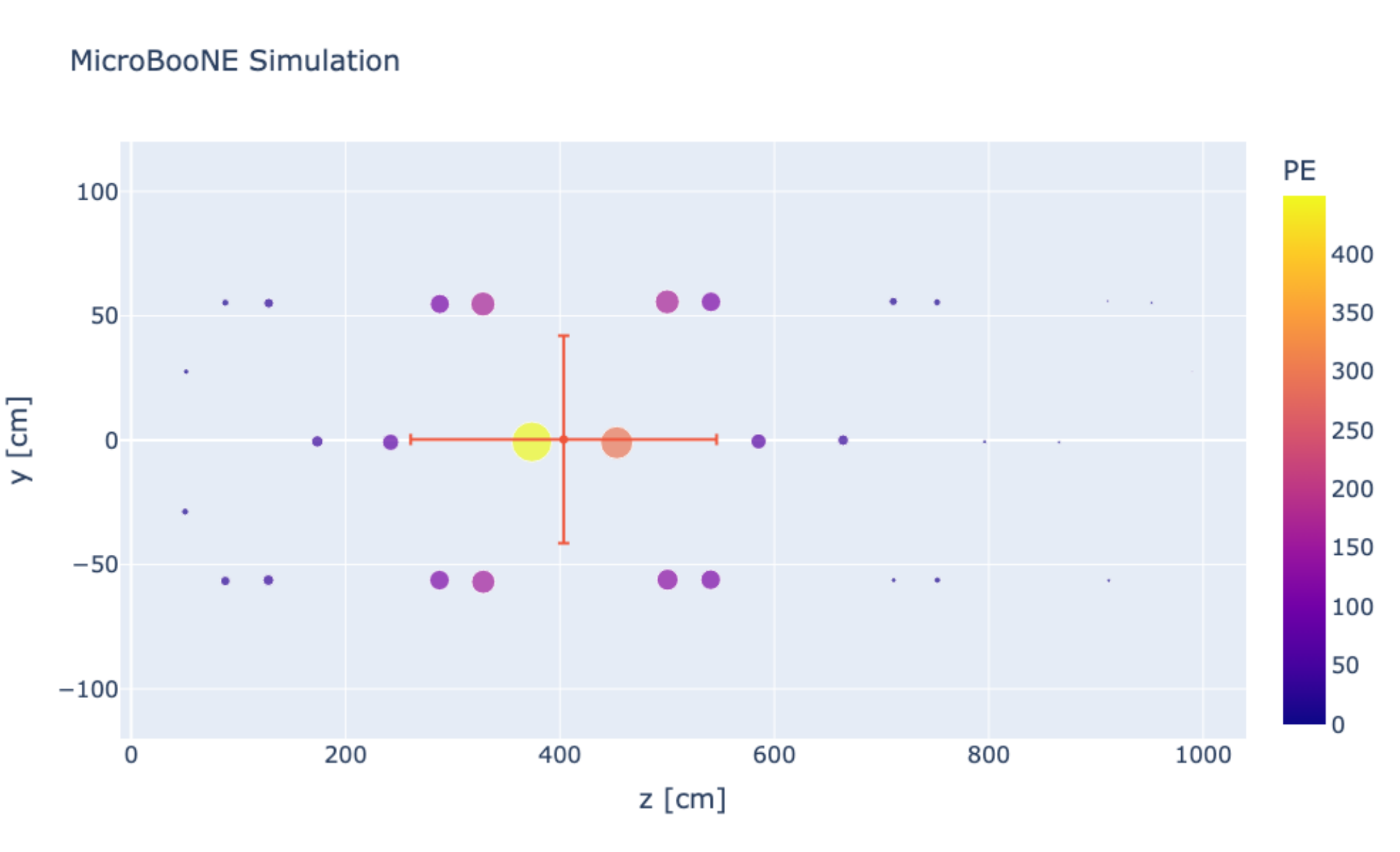}
   \caption{}
   \label{fig:optic_flash}
\end{subfigure}\\
\caption{Example of plots from the "Optical Information" notebook: the time of the optical hits with respect to the beam trigger (\subref{fig:optic_time}), the difference in wire units between the barycenter of the optical flash and the one of the neutrino hits as identified by Pandora (\subref{fig:optic_baryc}), display of the 32 PMT detectors with size and color weighted by the number of photoelectrons in a specific event, as well the flash barycenter and its width (\subref{fig:optic_flash}).}
\label{fig:optic}
\end{figure}

Finally, as the other notebooks focused on the LArTPC information, the "Optical Information" notebook demonstrates the usage of the optical detector information.
This notebook shows how to access optical hit properties, and how the MicroBooNE optical reconstruction clusters them in time into “flash” objects. It also shows how to use the light information to help identifying the neutrino hits in the TPC, e.g. by comparing the flash barycenter and with the one from the neutrino TPC hits. Examples are shown in figure~\ref{fig:optic}.

\section{Conclusions and Outlook}

MicroBooNE has released data sets for collaborative software development, and made them available on Zenodo and via xrootd. A wide range of applications can be developed based on these samples, with data formats for both image processing (e.g. CNN) and hit-based algorithms (GNN or other traditional algorithms). The size of the sample is enough for neural network training, and examples for extracting the ground truth for supervised learning are provided. The rich documentation for the usage of these data sets includes a set of performance metrics with reference results from the Pandora algorithms.

The statistics on the Zenodo website indicate that the data sets have been dowloaded hundreds of times already. The first developments based on these samples were presented at the conference.


%
\bibliography{bibliography}

\end{document}